\documentstyle[twoside,fleqn,espcrc2,epsfig]{article}
\newcommand{\ecm}{E_{\mathrm cm}}

\newcommand{\as}{$\alpha_s$}
\newcommand{\oas}{$\cal O$($\alpha_s^2$)}
\newcommand{\gev}{\mbox{Ge\kern-0.2exV}}
\newcommand{\mev}{\mbox{Me\kern-0.2exV}}

\newcommand{\DW}{Dokshitzer and Webber}
\newcommand{\asb}{$\bar{\alpha}_0$}
\newcommand{\beq}{\begin{equation}}
\newcommand{\eeq}{\end{equation}}
\newcommand{\bea}{\begin{eqnarray}}
\newcommand{\eea}{\end{eqnarray}}

\title{Determination of $\alpha_s$ from Event Shapes and Power Corrections}
\author{Daniel Wicke\address{Fachbereich Physik, 
         Bergische Universit\"at GH,
         Gau{\ss}str.~20,
         42097 Wuppertal;
         wicke@cern.ch
        }}
\begin{document}
\begin{abstract}
The size of non-perturbative corrections to event shape
observables is predicted to fall 
like a power of the inverse centre of mass energy. These power
corrections are investigated for different observables from
\mbox{e$^+$e$^-$}-an\-ni\-hi\-la\-tion and compared to the theoretical predictions.
Using the latest DELPHI high energy data advantages of determining
$\alpha_s$ from 
these predictions are discussed and compared to conventional methods.
\end{abstract}
\maketitle

\section{Introduction}
The upgrade of LEP to run at energies above the $Z$-peak induced new
possibilities in measuring
the running of \as. 
Due to asymptotic freedom, non-perturbative effects
become less important at higher energies. Different theoretical frameworks
suggest that for event shapes the non-perturbative corrections reduce like
the inverse power of the centre of mass energy
\cite{QCD97_Zakharov,PhysLettB352_451}.   
Theoretical predictions for this decrease can be used in the
measurement of \as\
independent of event by event hadronization models.

\section{$\protect\boldmath\alpha_s$ from Mean Event Shapes using Power Corrections}
The analytical  power ansatz for non-per\-tur\-ba\-tive corrections by \DW\ 
\cite{PhysLettB352_451,hep-ph/9510283}
can be  used to determine \as\ from mean event shapes
\cite{ZPhysC73_229,DELPHI97-92conf77,QCD97_Biebel}.
This ansatz provides an additive term to the perturbative \oas\ QCD
prediction $\left< f_{\mathrm pert} \right>$:
\beq
\left< f \right> = 
\frac{1}{\sigma_{\mathrm tot}}\int f\frac{df}{d\sigma}d\sigma =
\left< f_{\mathrm pert} \right> + \left< f_{\mathrm pow} \right> 
\label{eq_f}
\eeq
The power correction is given by
\[
\left < f_{\mathrm pow}\right >  =  a_f \cdot
\frac{\mu_I}{E_{\mathrm cm}}
  \Bigg[\bar{\alpha}_0(\mu_I) - \alpha_s(\mu)
\]\beq
       - \left(\beta_0 \cdot \ln{\frac{\mu^2}{\mu_I^2}} + 
\frac{K}{2\pi} + 2\beta_0 \right) \alpha_s^2(\mu) 
\Bigg]\quad{\rm .}
\label{eq_fpow_dw}
\eeq
\asb\ is a non-perturbative parameter accounting for the
contributions to the event shape below an infrared matching scale $\mu_I$,
$K=(67/18-\pi^2/6)C_A-5N_f/9$, $\beta_0=(33-2N_f)/12\pi$ and $a_f=4C_f/\pi$.
Beside \as\ this formulae contains \asb\ as the only free parameter. In
order to measure \as\ this  parameter has to be known.

To infer \asb\ a combined fit of \as\ and \asb\ to a large set of
measurements at different energies
\cite{ZPhysC73_229,DELPHI97-92conf77,collection_eventshapes} 
is performed. For $\ecm\ge M_Z$ only DELPHI measurements are included in
the fit. Fig.~\ref{fig_mess_fit} shows the measured mean values of
$\left<1-T\right>$ and $\left<M_h^2/E_{\mathrm vis}^2\right>$ as a function of
the centre of mass energy together with the results of the fit.
The resulting values of \asb\ are summarized in Tab.~\ref{tab_mess_fit}.
\begin{table}[b]
\caption{\label{tab_mess_fit}\asb\ as determined from mean event shapes.
}
\begin{center}
\begin{tabular}{|c|c|c|} \hline
Observable    & $\bar{\alpha}_0$  $\pm$ stat. $\pm$ scale         
              & $\chi^2/\mbox{ndf}$ \\ \hline
$\left<1-T\right>$ 
& $0.531 \pm 0.012 \pm 0.003$
& 42/23 \\
$\left<M_h^2/E_{\mathrm vis}^2\right>$ 
 & $0.434 \pm 0.010 \pm 0.010$
& 4.0/14 \\ \hline 
\end{tabular}
\end{center}
\end{table}
The extracted \asb\ values are around 0.5 as expected in 
\cite{hep-ph/9510283},
but they are incompatible with each other. As the assumed
universality is not given to the precision that is accessible from
data, \asb\ is individually determined for $\left<1-T\right>$ and
$\left<M_h^2/E_{\mathrm vis}^2\right>$. 
The scale error is obtained from varying the renormalization scale
$f=\mu^2/E_{\mathrm cm}^2$ from 0.25 to 4 and the infrared matching
scale from  $1 \gev$ to $2 \gev$.

After fixing \asb, the \as\ values corresponding to the high
energy data points can be calculated from 
Eqs.~(\ref{eq_f}--\ref{eq_fpow_dw}). \as\ is calculated for both
observables individually and then combined with an unweighted average.
Its error is propagated from the data and combined
by assuming full correlation. An additional scale error was calculated by 
varying $f$ and $\mu_I$ in the ranges discussed.
The results are summarized in Tab.~\ref{tab_as_mes} and plotted as
function of $\ecm$ (open dots) together with
the QCD expectation in Fig.~\ref{fig_as}a.
\begin{table}[tb]
\caption{\label{tab_as_mes}\as\ as obtained from means with the 
Eqs.~(\ref{eq_f}--\ref{eq_fpow_dw}) by averaging the $\left<1-T\right>$ and $\left<M_h^2/E_{\mathrm vis}^2\right>$ results.}
\begin{center}
\renewcommand{\arraystretch}{1.2}
\begin{tabular}{|c|c@{ $\pm$}c@{ $\pm$}c@{ $\pm$}c|}
\hline
$E_{\mathrm cm}$ &  $\alpha_s(E_{\mathrm cm})$ & stat. & sys. & scale \\
\hline      
133\gev          &  0.1162 & 0.0070 & 0.0009 & 0.0047 \\
161\gev          &  0.1047 & 0.0069 & 0.0034 & 0.0043 \\
172\gev          &  0.1052 & 0.0082 & 0.0037 & 0.0041 \\
\hline
\end{tabular}
\end{center}
\end{table}
\section{$\boldmath\alpha_s$ from Event Shapes Distributions}
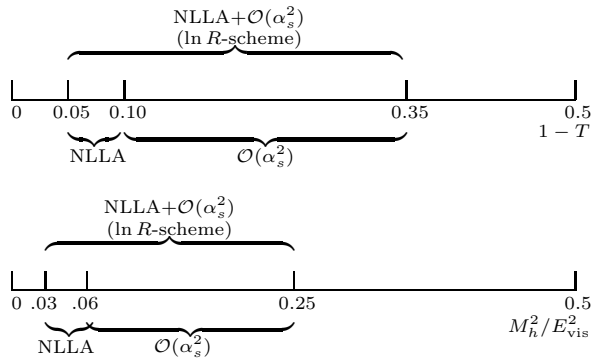
\begin{figure}
\scriptsize
\unitlength 0.5mm
\begin{picture}(150,50)
\put(  0,20){\line(1,0){150}}
\put(  0,20){\line(0,1){5}}
\put( 15,20){\line(0,1){5}}
\put( 30,20){\line(0,1){5}}
\put(105,20){\line(0,1){5}}
\put(150,20){\line(0,1){5}}
\put(0,15){0}
\put(11,15){0.05}
\put(26,15){0.10}
\put(101,15){0.35}
\put(147,15){0.5}
\put(140,10){$1-T$}
\put(15,30){$\overbrace{\makebox(89,0){}}$}
\put(15,12){$\underbrace{\makebox(14,0){}}$}
\put(30,12){$\underbrace{\makebox(75,0){}}$}
\put(16,42){\makebox(89,0){NLLA+${\cal O}(\alpha_s^2)$}}
\put(16,36){\makebox(89,0){($\ln R$-scheme)}}
\put(16,5){\makebox(14,0){NLLA}}
\put(30,5){\makebox(75,0){${\cal O}(\alpha_s^2)$}}
\end{picture}\\
\begin{picture}(150,50)
\put(  0,20){\line(1,0){150}}
\put(  0,20){\line(0,1){5}}
\put(  9,20){\line(0,1){5}}
\put( 20,20){\line(0,1){5}}
\put( 75,20){\line(0,1){5}}
\put(150,20){\line(0,1){5}}
\put(0,15){0}
\put( 5,15){.03}
\put(16,15){.06}
\put(71,15){0.25}
\put(147,15){0.5}
\put(132, 9){$M_h^2/E_{\mathrm vis}^2$}
\put( 9,30){$\overbrace{\makebox(66,0){}}$}
\put( 9,12){$\underbrace{\makebox(11,0){}}$}
\put(20,12){$\underbrace{\makebox(55,0){}}$}
\put( 9,42){\makebox(66,0){NLLA+${\cal O}(\alpha_s^2)$}}
\put( 9,36){\makebox(66,0){($\ln R$-scheme)}}
\put( 9,5){\makebox(11,0){NLLA}}
\put(18,5){\makebox(55,0){${\cal O}(\alpha_s^2)$}}
\end{picture}
\caption{\label{fig_fit_ranges}Fit ranges chosen for fitting \as\ from
         different QCD predictions of $1-T$ and $M_h^2/E_{\mathrm vis}^2$ 
         distributions. 
        }
\end{figure}

From event shape distributions \as\ is determined by fitting an \as\
dependent QCD 
prediction  (\oas, pure NLLA or combined \oas+NLLA
\cite{NuclPhysB178_412,PhysLettB263_491,PhysLettB272_368}) 
folded with a hadronization or a power correction to the data.
The fitranges used for the different QCD predictions are shown in
Fig.~\ref{fig_fit_ranges}. 
The ranges for pure NLLA and \oas\ fits are chosen to be distinct, so
that the results are statistically uncorrelated.

\subsection{Using Hadronization Models}
As an example of a Monte Carlo based model in this section
the hadronization correction is calculated using the JETSET PS model
(Version 7.4 tuned to DELPHI data \cite{ZPhysC73_11}). 
The QCD prediction is multiplied in each bin by the correction factor
(the model prediction of the ratio of hadron over parton level in that
bin).

For \oas-fits the renormalization scale $\mu$ was fitted to the LEP1 data
\cite{ZPhysC73_11} for both observables individually and then fixed for the fits
to the high energy distributions. This takes advantage of 
optimized scales.
In contrast for NLLA and combined fits $\mu$ was set equal to
$E_{\mathrm cm}$, so that these results can
be compared to other experiments.

The systematic errors were obtained by fitting several $1-T$ and
$M_h^2/E_{\mathrm vis}^2$ distributions evaluated by applying different cuts.
The mean deviation from the central value is used as
systematic error. Scale errors were taken from previous 
DELPHI publications~\protect\cite{ZPhysC54_55,ZPhysC59_21}.

\begin{figure*}[t]
\centerline{a)\hspace*{-0.8cm}\epsfig{file=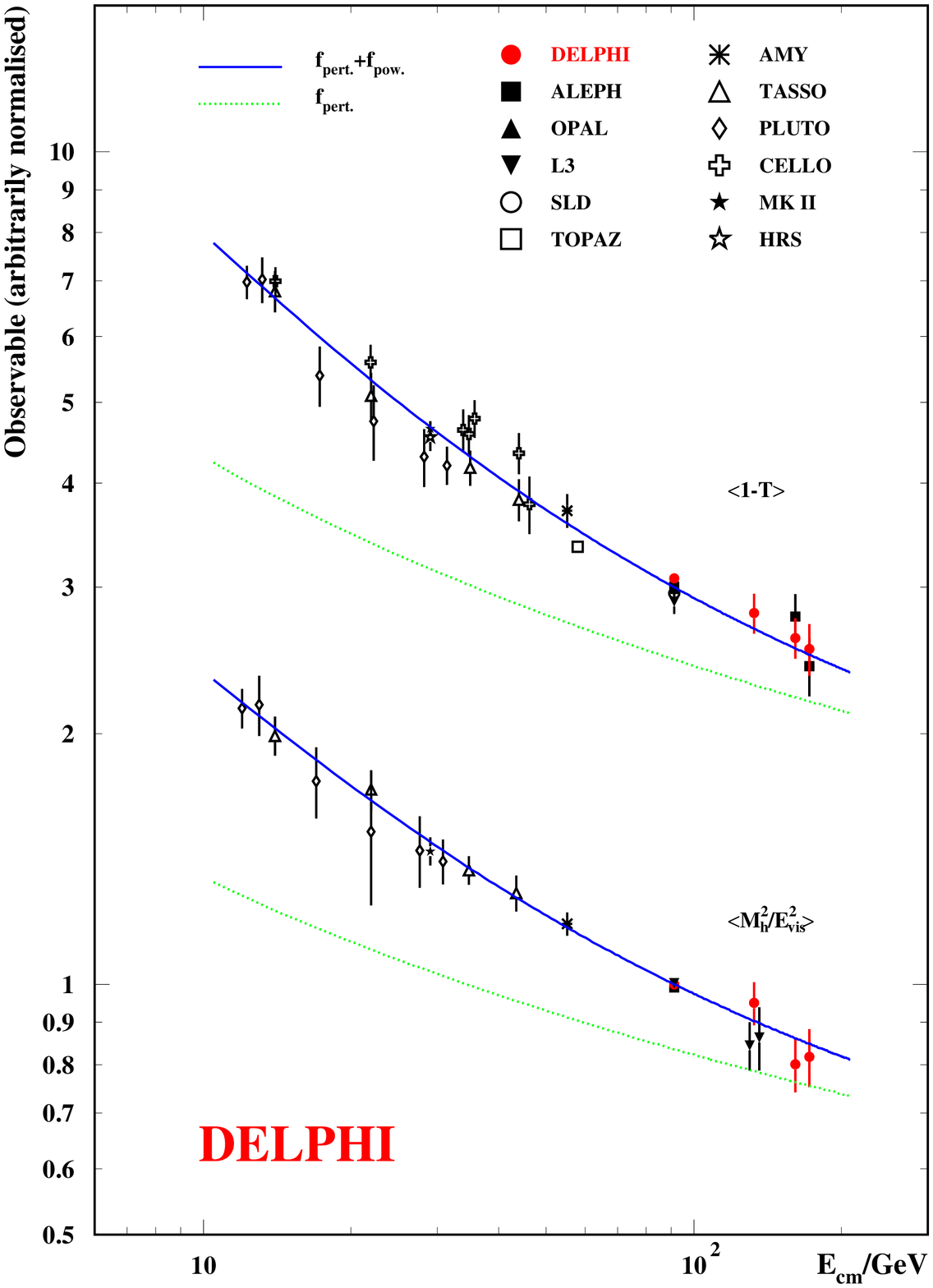,width=5.5cm}
        b)\hspace*{-0.8cm}\epsfig{file=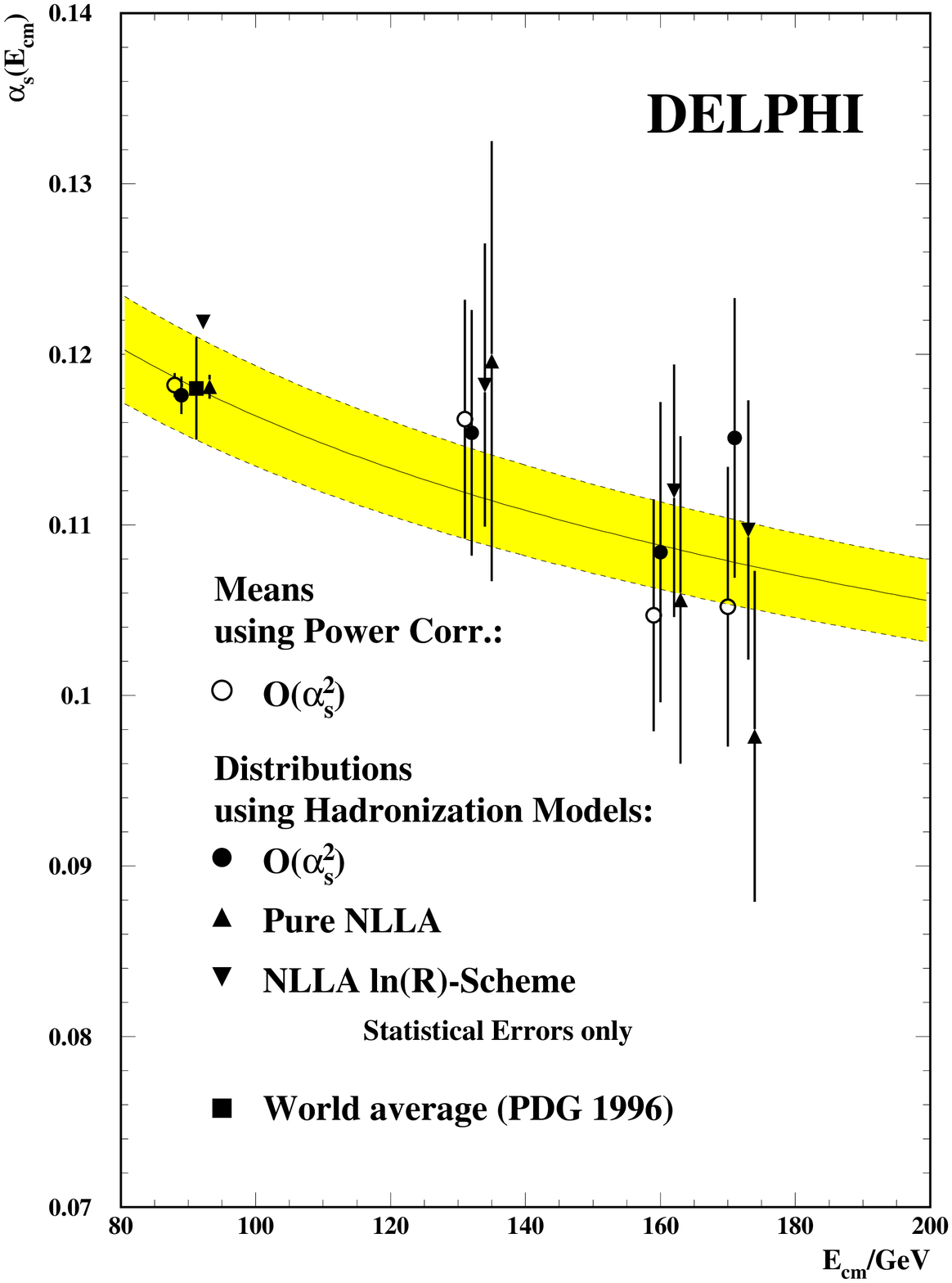,width=5.5cm}
        c)\hspace*{-0.8cm}\epsfig{file=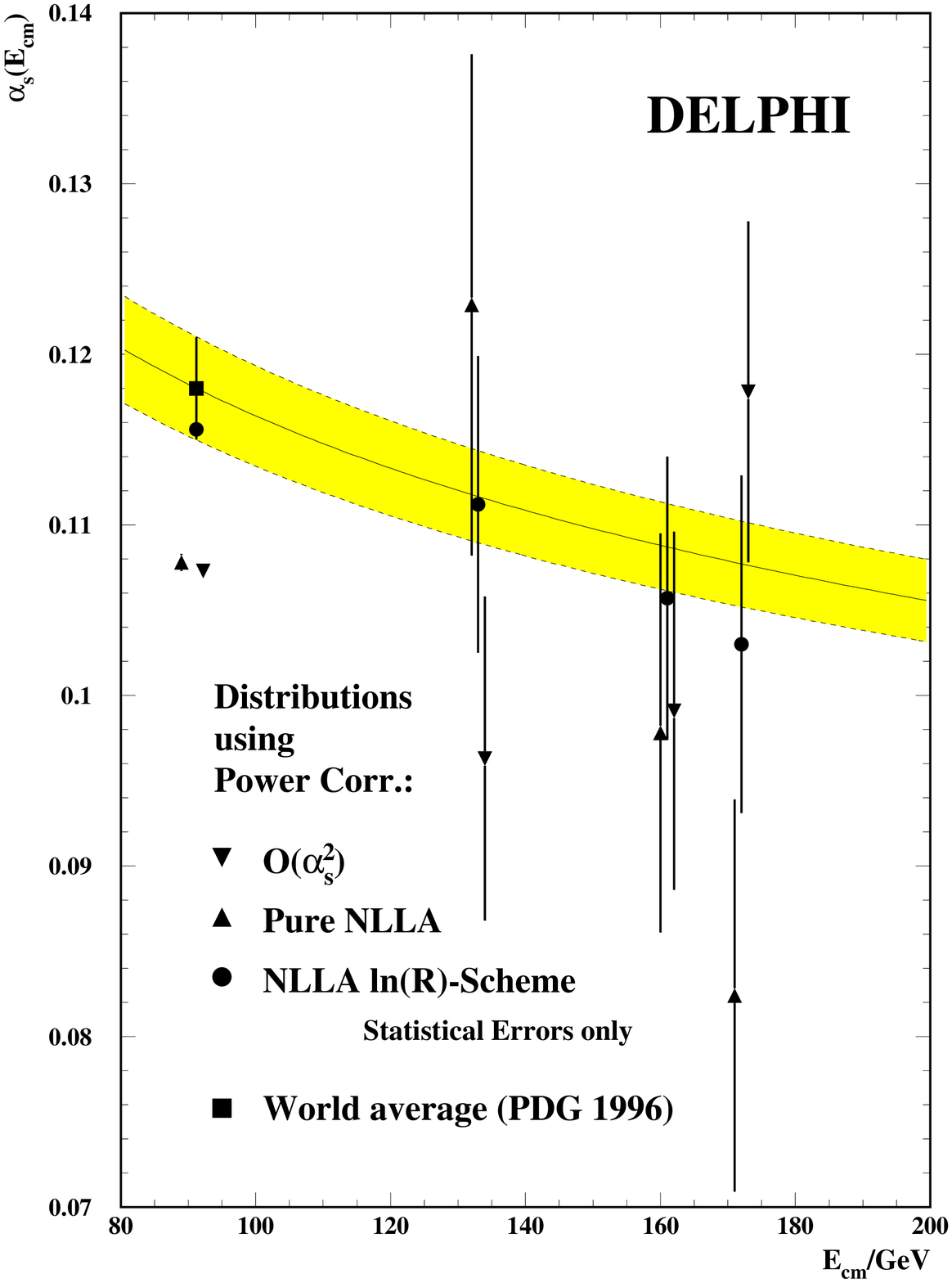,width=5.5cm}}
\caption{a)\label{fig_mess_fit}
Measured mean values of  $\left<1-T\right>$ and
$\left<M_h^2/E_{\mathrm vis}^2\right>$ as a function of the centre of mass
energy. The solid lines present the results of the fits with
Eqs.~(\protect\ref{eq_f}--\protect\ref{eq_fpow_dw}), the dashed lines show the
perturbative part.\protect\newline
b) \label{fig_as}Energy dependence of \as.
                     The errors shown are statistical
                     only. The band shows the QCD
                     expectation of extrapolating the world average to
                     other energies.\protect\newline
c) \label{fig_esd_dwas}\as\ determined using \DW\ power
corrections with $1-T$ distribution.}
\end{figure*}

The \as\ values evaluated from the distributions are given in
Tab.~\ref{tab_as_esd} and shown as a function of $\ecm$ in
Fig.~\ref{fig_as}b. The results agree well
with those measured from the event shape means. 

\subsection{Using Power Corrections}

\DW\ recently extended their model of power corrections to event shape
distributions \cite{hep-ph/9704298}. For distributions the power corrections are applied by
shifting the complete prediction:
\beq
F(f)=\frac{1}{\sigma_{\mathrm tot}}
     \frac{d\sigma}{df}=F_{\mathrm pert}(f-\delta f)
\label{eq_dw_esd}
\eeq
where the size of the shift is given by the size of the power
correction to the corresponding mean 
($\delta f = \left< f_{\mathrm pow} \right>$).
The perturbative prediction $F_{\mathrm pert}$ can now be any of the
available QCD predictions: \oas, NLLA or any combined scheme.

\begin{figure*}[p]
\begin{center}
\vspace*{-0.7cm}
\unitlength1mm
\nocite{DELPHI97-92conf77}
\mbox{  
        \begin{picture}(0,0)
        \put(61,80){a)}
        \put(36,85){$\chi^2/{\mathrm ndf}=120/107$}
        \end{picture}
        \epsfig{file=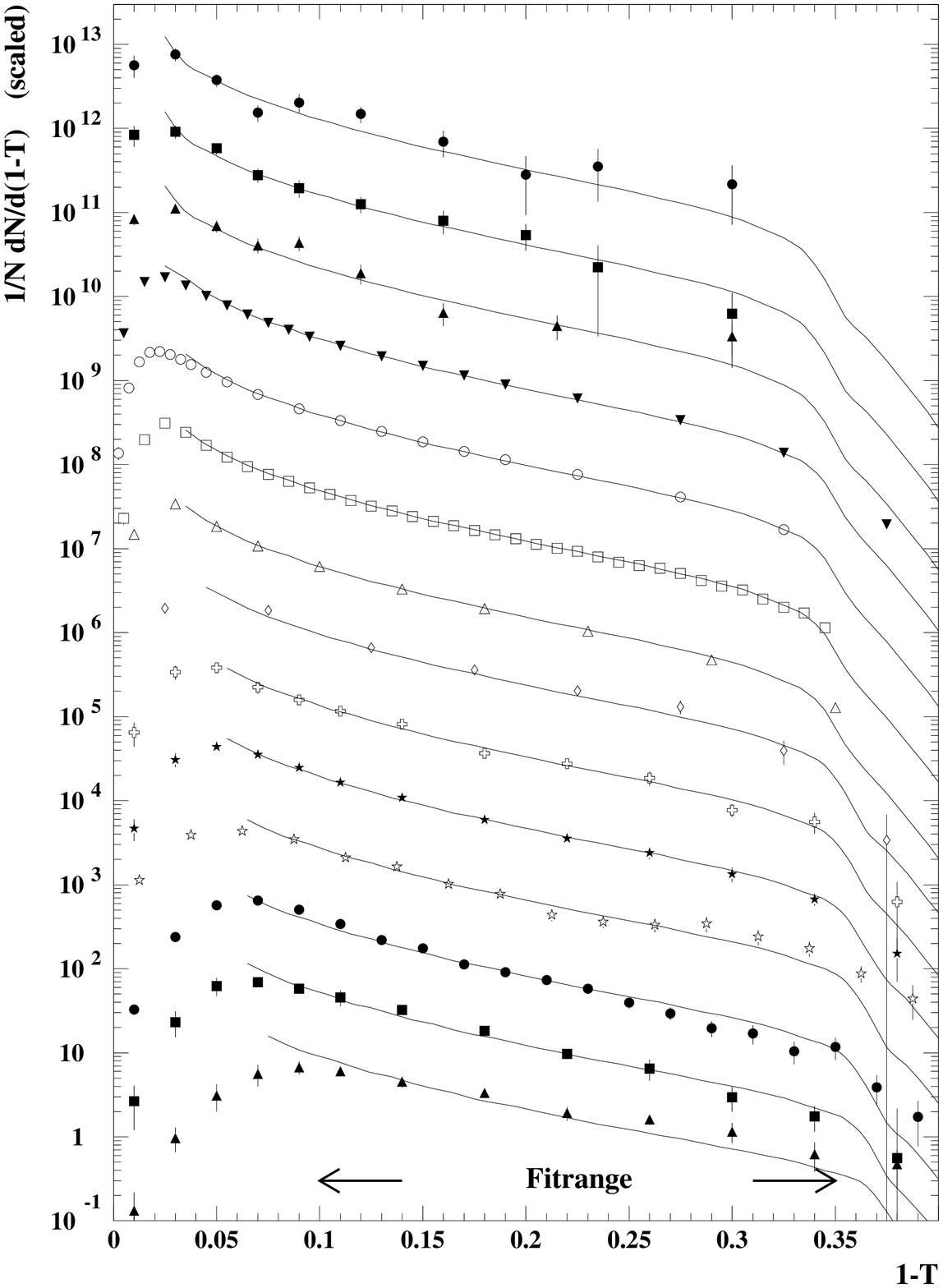,width=7.cm}
        \hspace*{-1.77cm}
        \begin{picture}(0,0)
        \put(61,80){b)}
        \put(36,85){$\chi^2/{\mathrm ndf}=416/108$}
        \end{picture}
        \epsfig{file=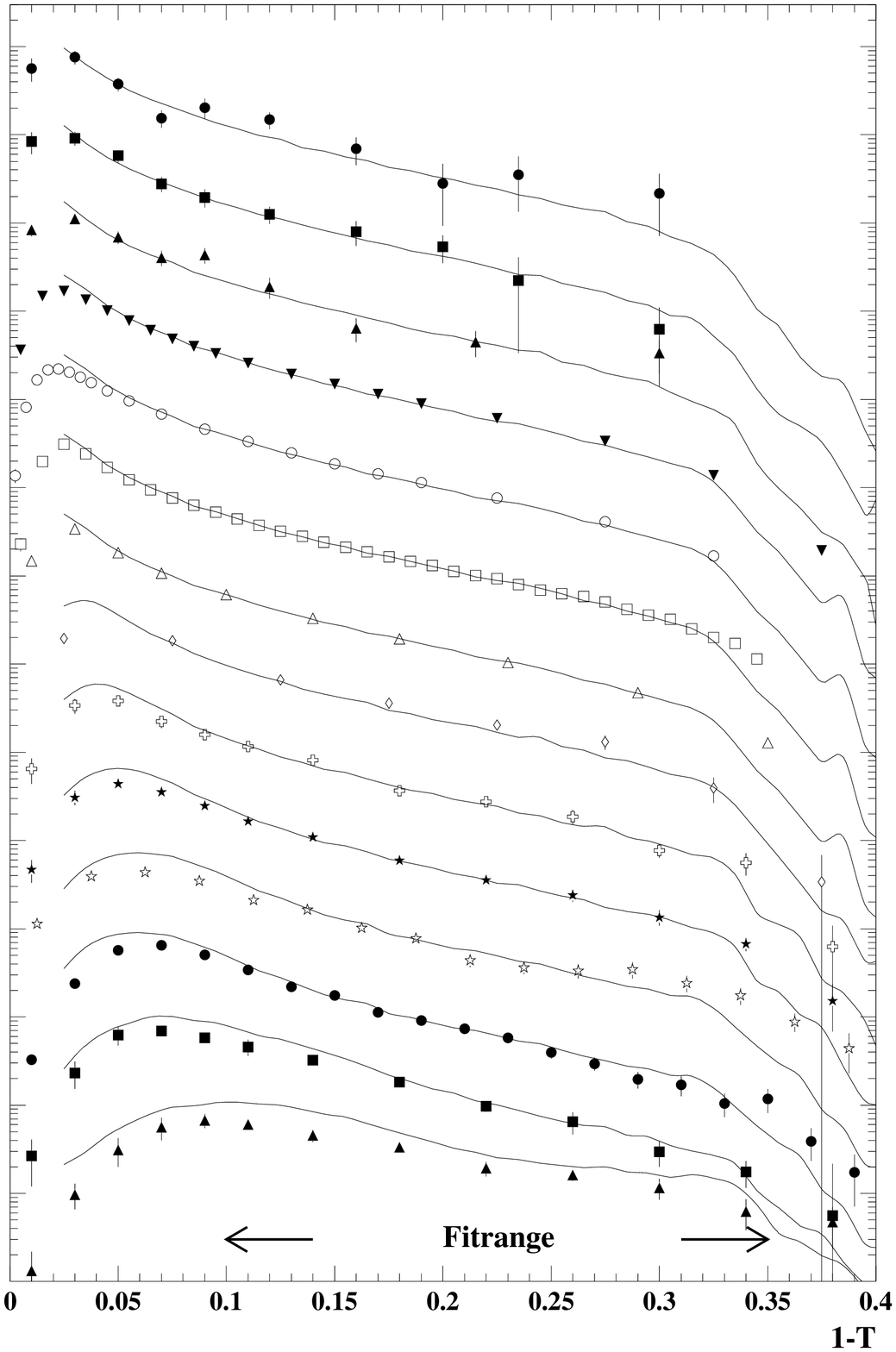,width=7.cm}
        \hspace*{-1.6cm}
        \epsfig{file=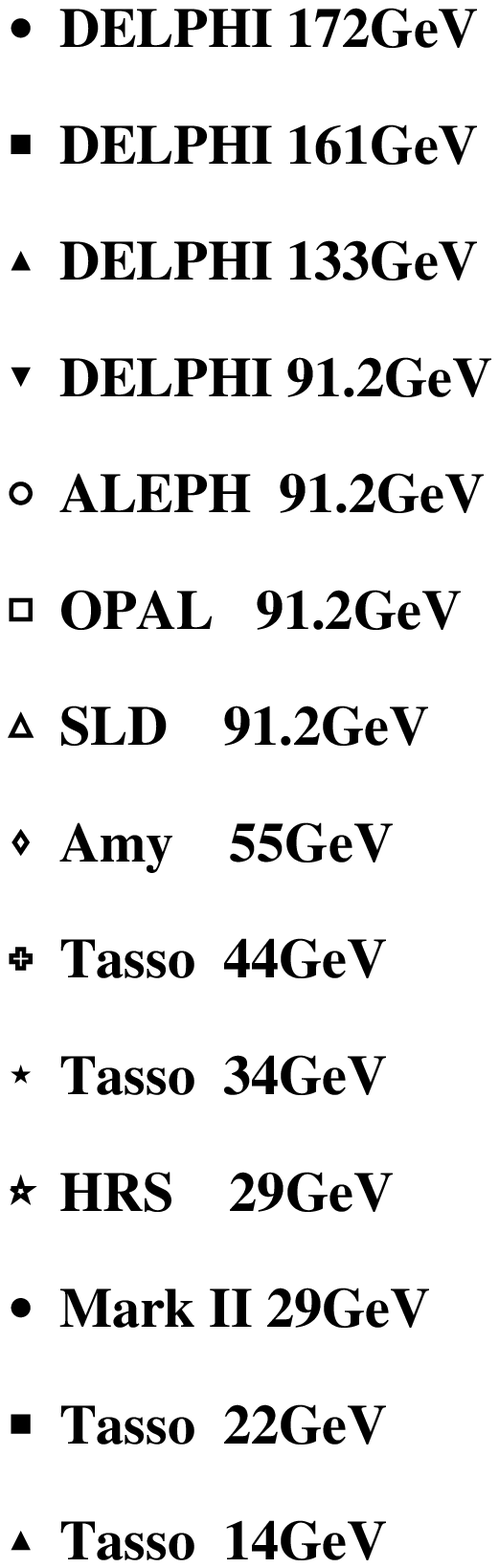,width=7.cm}
     }
\\\vspace*{-1cm}
\mbox{  
        \begin{picture}(0,0)
        \put(61,80){c)}
        \put(39,85){$\chi^2/{\mathrm ndf}=83/48$}
        \put(27,0){Power Correction}
        \end{picture}
        \epsfig{file=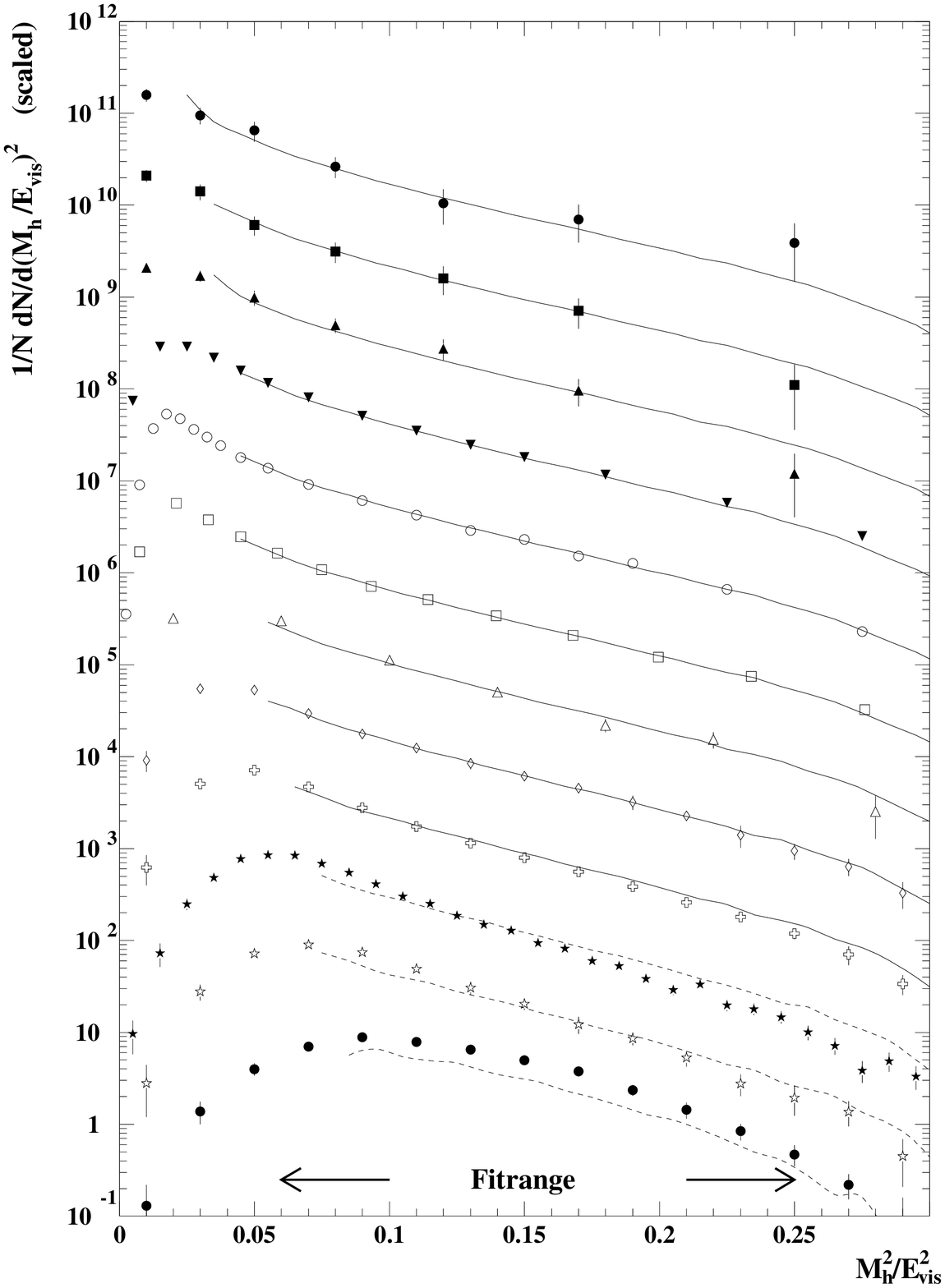,width=7.cm}
        \hspace*{-1.77cm}
        \begin{picture}(0,0)
        \put(61,80){d)}
        \put(39,85){$\chi^2/{\mathrm ndf}=55/49$}
        \put(35,0){JETSET}
        \end{picture}
        \epsfig{file=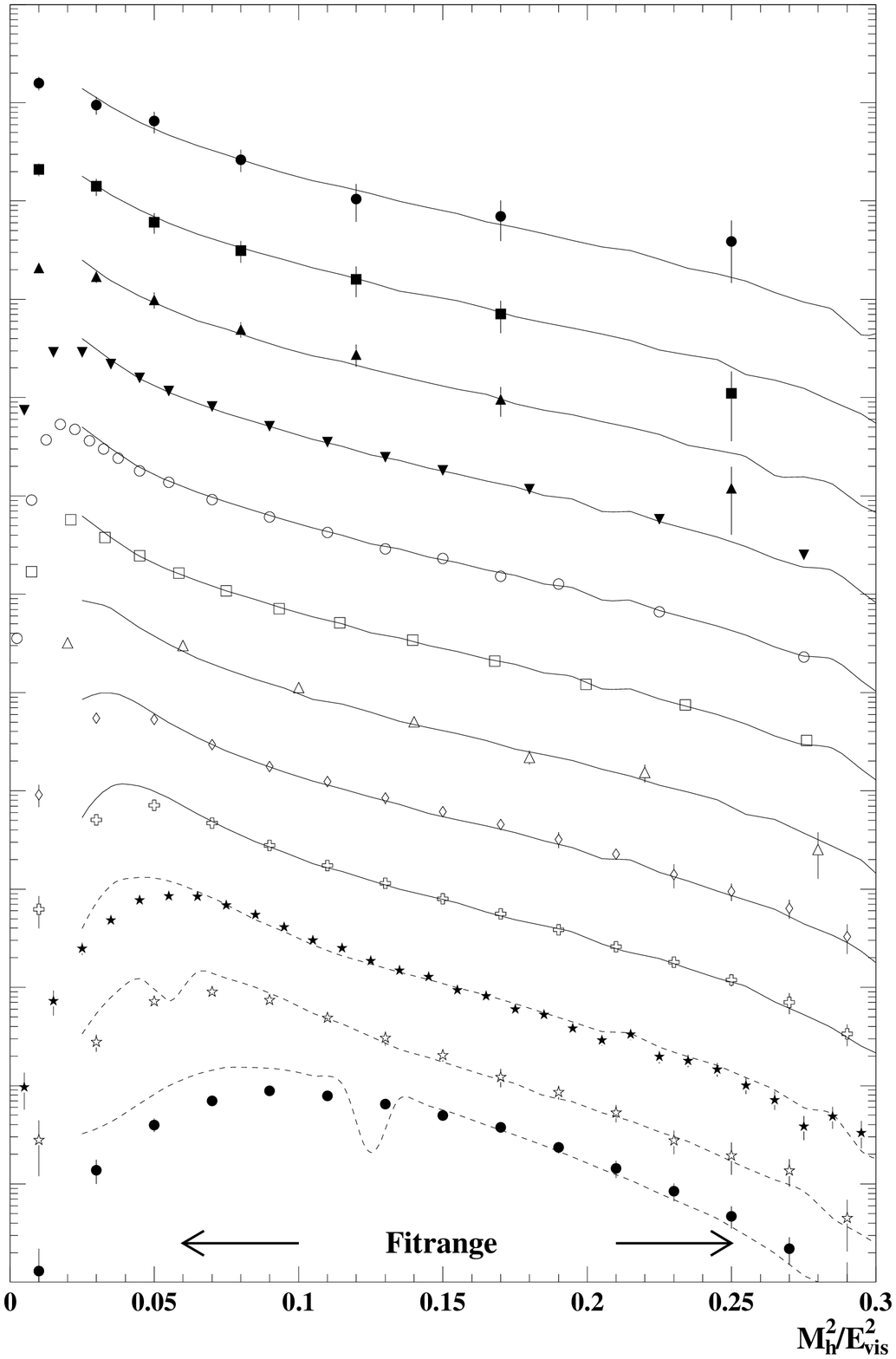,width=7.cm}
        \hspace*{-1.6cm}
        \epsfig{file=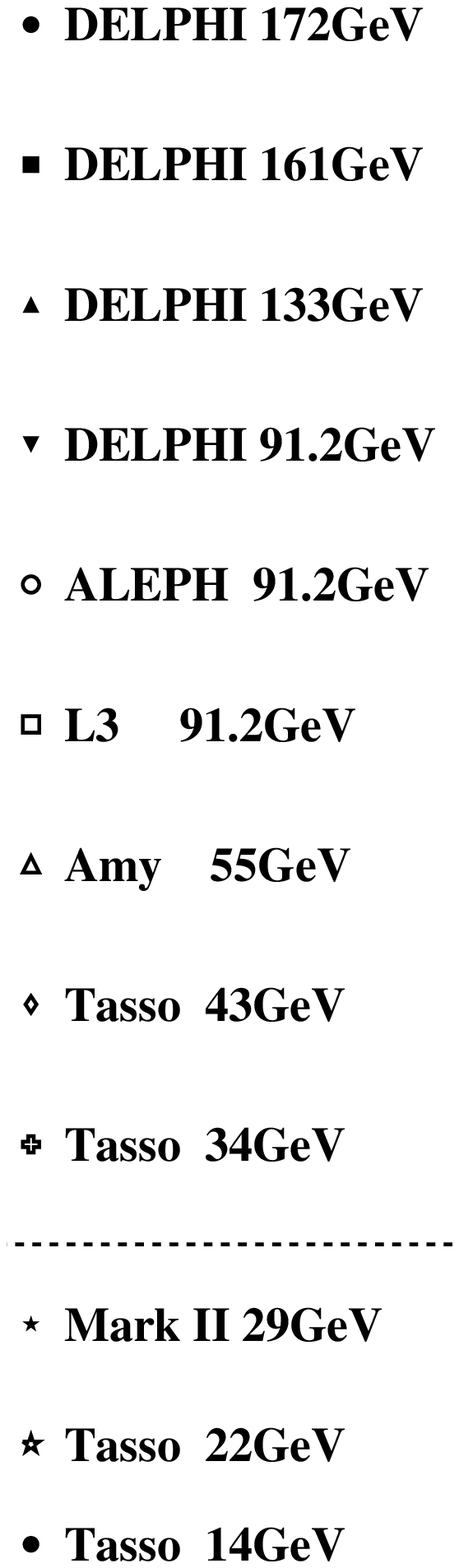,width=7.cm}
     }
\end{center}
\vspace*{-0.7cm}
\caption{\label{fig_esd_fit}\oas\ fit of \as\ (and \asb) to $1-T$ and
$M_h^2/E_{\mathrm vis}^2$  distributions  
measured at different energies. 
In plot a,c the power ansatz Eq.~\protect\ref{eq_dw_esd}, in b,d
JETSET~PS based correction is used to include
non-perturbative contributions. The distributions including dotted lines 
were not included in the fit. 
}
\end{figure*}

To measure \as\ at a single energy, again one first has to
determine \asb. This is done in a combined fit of \asb\ and \as\ to
measurements \cite{ZPhysC73_229,DELPHI97-92conf77,collection_eventshapes}
at many different energies.
See Fig.~\ref{fig_esd_fit}a,~c and Tab.~\ref{tab_esd_t}. 
For $M_h^2/E_{\mathrm vis}^2$
some of the lowest energies were removed from the fit, because the
shifted QCD prediction was no longer defined at the lower end of the fit
range.

To compare the quality of these fits to a hadronization model based
description, a simultaneous fit of \as\
to these data using the JETSET PS based hadronization correction is
done. The results are plotted in Fig.~\ref{fig_esd_fit}b,~d and summarized in
Tab.~\ref{tab_esd_t}.

Using these \asb\ results to determine \as\ from individual high
energy $1-T$ distributions lead to systematic deviations from the QCD
expectation (See Fig.~\ref{fig_esd_dwas}c). 
The deviation is different for NLLA and \oas\ (which use different
fit ranges) and opposite in sign. 

To investigate this problem, the \DW\ power correction term was
replaced by a simple ansatz:
\beq
\delta f = \frac{C_1}{E_{\mathrm cm}} +\frac{C_2}{E^2_{\mathrm cm}}
\eeq
A combined fit of $C_1$, $C_2$ to the complete set of
distributions probes the need of the quadratic power correction
term. \as\ was fixed to 0.120 here. 
It is found, that for the fit ranges used in \oas\ and NLLA fits of Thrust
the correction arising from the quadratic power term is not
negligable (See Tab. \ref{tab_esd_simplefit}). The small result of
$C_2$ for the combined result thus may be accidental.

\begin{table}[p]
\caption{\label{tab_as_esd}\as\ as obtained from  distributions using
JETSET.
}
\begin{center}
\renewcommand{\arraystretch}{1.2}
\begin{tabular}{|c|c@{$\pm$}c@{$\pm$}c@{$\pm$}c|}
\hline
133\gev  & $\alpha_s(E_{\mathrm cm})$ & \multicolumn{2}{c@{$\pm$}}{exp.} & scale \\
\hline
        \oas             &0.1154&\multicolumn{2}{c@{$\pm$}}{0.0072} & 0.006\\
         NLLA            &0.1196&\multicolumn{2}{c@{$\pm$}}{0.0129} & 0.005\\
        Combined         &0.1182&\multicolumn{2}{c@{$\pm$}}{0.0083} & 0.006\\
\hline                                            
\hline                                           
161\gev  & $\alpha_s(E_{\mathrm cm})$ & stat. & sys. & scale \\
\hline                                            
        \oas             &0.1084&0.0088&0.0049 & 0.006\\
        NLLA             &0.1056&0.0096&0.0043 & 0.005\\
        Combined         &0.1120&0.0074&0.0032 & 0.006\\
\hline                                           
\hline                                           
172\gev  & $\alpha_s(E_{\mathrm cm})$ & stat. & sys. & scale \\
\hline                                           
        \oas             &0.1151&0.0082&0.0058 & 0.005\\
        NLLA             &0.0976&0.0097&0.0046 & 0.005\\
        Combined         &0.1097&0.0076&0.0029 & 0.005\\
\hline
\end{tabular}
\end{center}
\end{table}
\begin{table}[p]
\caption{\label{tab_esd_t}\asb\ as determined from event shape distributions.
The error is the statistical error from the fit.}
\begin{center}
\begin{tabular}{|c|c|c|} \hline
$1-T$         & $\bar{\alpha}_0$  & $\chi^2/\mbox{ndf}$ \\
 \hline
\oas     & $0.550 \pm 0.012$& 120/107 \\
NLLA     & $0.475 \pm 0.006$&  81/26 \\
Combined & $0.491 \pm 0.005$& 377/137 \\
\hline 
\hline
$M_h^2/E_{\mathrm vis}^2$ & $\bar{\alpha}_0$  & $\chi^2/\mbox{ndf}$ \\
 \hline
\oas     & $0.698 \pm 0.014$&  83/49 \\
NLLA     & $0.480 \pm 0.012$&  3.2/7 \\
Combined & $0.554 \pm 0.005$& 285/53 \\
\hline 
\end{tabular}
\end{center}
\end{table}
\begin{table}[p]
\caption{\label{tab_esd_simplefit}Results of fitting the simple power
ansatz to event shape distributions. \as=0.120.} 
\begin{center}
\begin{tabular}{|c|c|l|}
\hline
       $1-T$ &  $C_1 [\gev]$   & $C_2 [\gev^2]$\\
\hline            
      \oas    &  $1.23\pm 0.09$ & $ -10.4~\pm 1.3 $\\
       NLLA    &  $0.27\pm 0.02$ & $+10.0~\pm 0.6 $\\
    Combined & $0.73\pm 0.02$ & $-~0.48\pm 0.61 $\\
\hline            
\hline            
$M_h^2/E_{vis}^2$  &  $C_1 [\gev]$   & $C_2 [\gev^2]$\\
\hline            
       \oas    &  $0.27\pm 0.09$ & $+14.8~\pm 6.9 $\\
       NLLA    & \multicolumn{2}{c|}{not converging}\\
       Combined&  $0.70\pm 0.02$ & $-~4.5~\pm 6.9 $\\
\hline
\end{tabular}
\end{center}
\end{table}

\section{Summary/Conclusions}
Several methods of  determining \as\ at high energies from event
shapes were discussed. 

The results from means using power corrections are compatible with
those from distributions using standard hadronization corrections.
Results from power corrections yield smaller scale errors.

Power corrections to event shape distributions are more tricky.

\newcommand{\esdcollection}
{   
 ALEPH  Coll., {\em Phys. Lett.} {\bf B284} (1992) 163.
\\   
 ALEPH Coll., {\em Z. Phys.} {\bf C55} (1992) 209.
\\   
 AMY  Coll., {\em Phys. Rev. Lett.} {\bf 62} (1989) 1713.
\\   
 AMY Coll., {\em Phys. Rev.} {\bf D41} (1990) 2675. 
\\   
 CELLO Coll., {\em Z. Phys.} {\bf C44} (1989) 63. 
\\   
 HRS Coll., {\em Phys. Rev.} {\bf D31} (1985) 1.
\\   
 JADE Coll., {\em Z. Phys.} {\bf C25} (1984) 231.
\\   
 JADE Coll., {\em Z. Phys.} {\bf C33} (1986) 23.
\\   
 L3 Coll., {\em Z. Phys.} {\bf C55} (1992) 39.
\\   
 Mark II Coll., {\em Phys. Rev.} {\bf D37} (1988) 1. 
\\   
 Mark II Coll., {\em Z. Phys.} {\bf C43} (1989) 325.
\\   
 MARK J Coll., {\em Phys. Rev. Lett.} {\bf 43} (1979).
\\   
 OPAL Coll., {\em  Z. Phys.} {\bf C59} (1993) 1.
\\   
 PLUTO Coll., {\em Z. Phys.} {\bf C12} (1982) 297. 
\\   
 SLD Coll., {\em Phys. Rev.} {\bf D51} (1995) 962.
\\   
 TASSO Coll., {\em Phys. Lett.} {\bf B214} (1988) 293.
\\   
 TASSO Coll., {\em Z. Phys.} {\bf C45} (1989) 11.
\\   
 TASSO Coll., {\em Z. Phys.} {\bf C47} (1990) 187. 
\\   
 TOPAZ Coll.,  {\em Phys. Lett.} {\bf B227} (1989) 495.
\\   
 TOPAZ Coll., {\em Phys. Lett.} {\bf B278} (1992) 506. 
\\   
 TOPAZ Coll., {\em Phys. Lett.} {\bf B313} (1993) 475}

\bibliographystyle{unsrtnewnt}
\bibliography{QCD}

\begin{thebibliography}{10}

\bibitem{QCD97_Zakharov}
V.~I. Zakharov.
\newblock In these proceedings.

\bibitem{PhysLettB352_451}
Y.~L. Dokshitzer and B.~R. Webber.
\newblock {\em Phys. Lett.} {\bf B352}(1995)  451.

\bibitem{hep-ph/9510283}
B.~R. Webber.
\newblock Talk given at workshop on DIS and QCD in Paris, hep-ph/9510283, 1995.

\bibitem{ZPhysC73_229}
{DELPHI Coll., P. Abreu et al.}
\newblock {\em Z. Phys.} {\bf C573}(1997)  229.

\bibitem{DELPHI97-92conf77}
J.~Drees, A.~Grefrath, K.~Hamacher, O.~Passon, and D.~Wicke.
\newblock DELPHI 97-92 CONF 77, contrib. to EPS HEPC  in
  Jerusalem, 1997.

\bibitem{QCD97_Biebel}
O.~Biebel.
\newblock In these proceedings.

\bibitem{collection_eventshapes}
\esdcollection.

\bibitem{NuclPhysB178_412}
R.~K. Ellis, D.~A. Ross, and A.~E. Terrano.
\newblock {\em Nucl. Phys} {\bf B178}(1981)  412.

\bibitem{PhysLettB263_491}
S.~Catani, G.~Turnock, B.~R. Webber, and L.~Trentadue.
\newblock {\em Phys. Lett.} {\bf B263}(1991)  491.

\bibitem{PhysLettB272_368}
S.~Catani, G.~Turnock, and B.~R. Webber.
\newblock {\em Phys. Lett.} {\bf B272}(1991)  368.

\bibitem{ZPhysC73_11}
{DELPHI Coll., P. Abreu et al.}
\newblock {\em Z. Phys.} {\bf C73}(1996)  11.

\bibitem{ZPhysC54_55}
{DELPHI Coll., P. Abreu et al.}
\newblock {\em Z. Phys.} {\bf C54}(1992)  21.

\bibitem{ZPhysC59_21}
{DELPHI Coll., P. Abreu et al.}
\newblock {\em Z. Phys.} {\bf C59}(1993)  21.

\bibitem{hep-ph/9704298}
Y.~L. Dokshitzer and B.~R. Webber.
\newblock Cavendish-HEP-97/2, hep-ph/9704298, 1997.

\end{thebibliography}
\end{document}